\documentclass{article}
\usepackage{jfrExamplee}
\usepackage{graphicx}
\usepackage{biblatex}
\usepackage{setspace}
\usepackage{subcaption}

\usepackage{amsmath}
\usepackage{url}

\title{Decentralized Multi-agent Filtering}

\author{
Dom Huh \\
University of California, Davis\\
Davis, CA 95616 \\
\texttt{dhuh@ucdavis.edu} \\
\And
Prasant Mohapatra \\
University of California, Davis\\
Davis, CA 95616 \\
\texttt{pmohapatra@ucdavis.edu} \\
}

\begin{document}

\maketitle

\begin{abstract}
This paper addresses the considerations that comes along with adopting decentralized communication for multi-agent localization applications in discrete state spaces. In this framework, we extend the original formulation of the Bayes filter, a foundational probabilistic tool for discrete state estimation, by appending a step of greedy belief sharing as a method to propagate information and improve local estimates' posteriors. We apply our work in a model-based multi-agent grid-world setting, where each agent maintains a belief distribution for every agents' state. Our results affirm the utility of our proposed extensions for decentralized collaborative tasks. The code base for this work is available in the following repo\footnote{\url{https://github.com/domhuh/ma-gym}}.
\end{abstract}

\section{Introduction}
Decentralization of multi-agent systems gives rise to modular joint behavior, as such systems constrain agents to observe, act and communicate locally. With this appeal of the increased flexibility, practicality, and survivability \cite{DurrantWhyte2006ABG}, there exists a rich history of literature surveying techniques on diverse and complex Dec-MAS applications \cite{240381, beard2020solution, afshari2021multi}. In more recent years, such works have largely focused on leveraging such methods for control tasks \cite{10.5555/1642293.1642498, Dames2015DetectingLA, 9369190, DBLP:journals/corr/abs-2101-05436, zhang2020lyapunov}. Our efforts in this work largely concentrate on developing a rich and well-defined feature space for such control applications while still adhering to decentralization constraints.

To realize the utility of decentralized multi-agent systems (Dec-MAS), we address some considerations that arise from the nature of decentralized communication. At its core, decentralized networks strictly limit communication on a peer-to-peer basis, hence only establishing links grounded on physical realities (i.e. sensing range). Thereby, agents often rely on a form of message-passing instead of using centralized communication facilities to transfer information between agents. Prior works \cite{GRIME1994849} have remarked such a restriction would result in inconsistencies in knowledge bases, unless a fully-connected topology is employed, which is often intractable. This inconsistency would require greater measures for proper coordination and limited predictability. Our proposed extension to the Bayes filtering algorithm attempts to mitigate these issues without violation of the locality constraint.

In this paper, we present a study on extending decentralization concepts to a foundational non-parametric state estimation algorithm, the Bayes filter \cite{KAELBLING199899}, on partially-observable tracking applications. This work adopts many ideas from prior works \cite{10.1145/1082473.1082522, NIPS2008_6c3cf77d, doi:10.1177/0278364920957090}, but we concentrate on the task of collaborative tracking to evaluate our proposed approach of greedy belief sharing for probabilistic state estimation.

\begin{figure}
    \centering
    \includegraphics[width=0.9\linewidth]{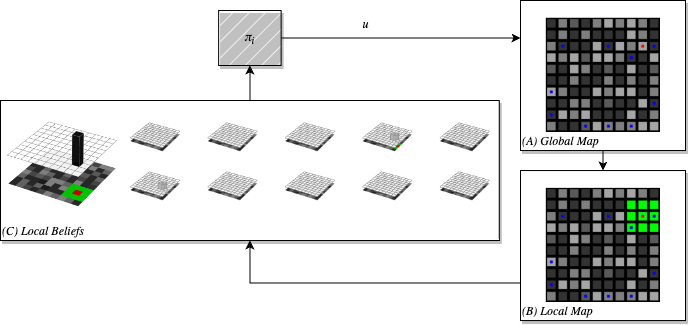}
    \caption{Dec-MAS Filtering Overview: (A) The dynamics and latent state of the environment is processed at the global map level. (B) Aspects of the global information are masked to preserve the decentralization property to construct the local information. (C) Each agent utilizes the local information provided from (B) to update and process their internal beliefs, estimating the state of every agent, to make informed decisions. }
    \label{fig:system_overview}
\end{figure}

\section{System Overview}\label{system_overview}
A general diagram of our Dec-MAS filtering system applied for localization is shown in Fig \ref{fig:system_overview}. As the task we choose to approach largely influences how each system will be formulated, we describe the task in this section. In our application, each agent $a_i \in A$, where $A$ is the set of all agents, maintains its own local beliefs $B_i$ using information coming from the local map level $L_i$. To reiterate, the objective of this system is not devising an novel control policy $\pi$, but rather we seek to develop a dynamic feature space that a control policy can leverage for various Dec-MAS control tasks. 

\subsection{Task Formulation.}
We focus our efforts on Dec-MAS, and follow the framework of Dec-POMDP. More specifically, we work in a 2-dimensional $H\times W$ grid world, where multiple agents reside. Each agent can only access their own local observation, taken from the cell they occupy. Additionally, the agents may access the observations and controls of the cooperative agents within its sensing range. If there are external entities within the sensing range, an indicator bit is appended to the agent's observation but no other information as we assume they are not cooperative. A more detailed sensing observation may be appended, but is not applied in this work.

The model of the environment is known to every agent (i.e. the transition dynamics and the emission probability), and is defined to be stationary, deterministic and surjective (i.e. aliased cell features). The transition dynamics $\mathrm{T}$ follows the king-graph, built upon the agent's control space $U$ consisting of Chebyshev movements. The sensing range is enforced by a threshold on the same distance measure. An ordering is enforced, avoiding any ties that may occur when multiple agents try to occupy the same cell. The emission probability defines the mapping from local observations to a subset of candidates in the latent state space. We extend the formalism of this Dec-MAS to the following tasks: Congregation and Predator-Prey. The finite time horizons for both tasks are discrete and limited to $t_{done}=30$.

\paragraph{Congregation.} The objective of the task is to have as many agents within each other's sensing range, hence the reward function is defined to be the sum of the total number of agents within each agent's sensing range. For this task, all agent follow the same control policy $\pi_i: B_i \rightarrow U$, mapping from a local belief space to the control space, and moving each agent closer to the others. Hence, this policy computes the likely positions of every agents using their own belief, breaking ties through random sampling. The action is chosen to move the agent towards the direction of the average likely position of all agents.

\paragraph{Predator-Prey.} The objective of the task is to have a group of agents, called the predators, find and catch another group of agents, called the preys. The reward function is the total number of predators that have a prey within their sensing range. For the predators, we move the agent toward where it believes the prey is at while also maintaining separation between each other. The prey's policy uses the same policy defined for Congregation, but instead of moving towards the average position of the predators, the prey moves away from it. In our experiments, the preys operate on global information. 

For both tasks, we establish a joint behavior that is stationary, and do not make any further regards towards devising any methods for greater planning, stability, nor adaptability.

\subsection{Decentralized Bayes Filter}
Each agent will locally maintain a state estimation for every agent. To compute the agent's belief, we use the standard Bayes filter algorithm, updating on local controls and observations. To incorporate the messages from nearby agents, we greedily update the agent's beliefs using neighboring beliefs. Hence, we perform the following recursive updates: control, observation, and belief sharing. For external entities (i.e. preys), we append their respective state estimation, following the same updates rules with minor modifications. All belief priors are initialized to be uniformly distributed over the latent state space.

\paragraph{Control update.} In this initial step, we use the controls of the agent to update the local posterior belief using the transition dynamics and prior belief. For discrete settings, we can map the transition dynamics $p(x'|x,u)$ as a linear tensor operator $\mathcal{T}_t$ that accounts for the controls taken at time-step $t$.
\begin{equation}
\mathcal{T}_{t} = \begin{bmatrix}
                    p(X|X,u_{0,t})\\
                    \dots\\
                    p(X|X,u_{n,t})
                  \end{bmatrix}
\end{equation}
where $n$ is the number of agents, $X$ denotes the latent state space, and thereby $p(X|X,u_{i,t})$ signifies the transition matrix over $X$ for agent $i$ if it takes the action $u_{i,t}$. Thereby, we can formulate our update in the following manner.
\begin{equation}
\bar{\textit{bel}}_{t+1} = \mathcal{T}_{t} \textit{bel}_{t}
\end{equation}
For agents outside of each other's sensing range, we mask the observed control of that agent within the transition dynamics, allocating for a masked action $u_m$, and set its transition probability to be the expectation over all possible controls $U' = U\backslash u_m$.
\begin{equation}
p(x'|x,u_m) = \frac{1}{|U'|} \sum_{u\in U'} p(x'|x,u)
\end{equation}
\paragraph{Observation update.} We then process the observations $z_i$ of each local agent, filtering the posterior of each belief based on the emission probability $p(z_i|x)$.
\begin{equation}
\textit{bel}_{i,t+1} = \eta (p(z_i|x) \odot \bar{\textit{bel}}_{i,t+1})
\end{equation}
where $\eta$ is a normalization constant so that the filtered posterior integrates to $1$. Again, for agents outside of the sensing range, we allocate a masked observation within the emission probability, defined to be uniform across the grid world.

\paragraph{Belief sharing update.} Once the posteriors are filtered, the agents collaboratively improve their local estimations through a fixed number of message passing rounds. With these multiple rounds, we introduce transitivity and a wider implicit communication range. Specifically, each agent $i$ and the set of neighboring agents $N_i$ communicate their beliefs, and greedily update its posteriors based on an entropy measure.
\begin{equation}
\textit{bel}_{(i,j),t} = \mathop{\text{argmin}}_{j\in A} \mathcal{H}(\textit{bel}_{(n,j),t}), n\in N_i
\end{equation}
We assert that such greedy updates will only monotonically improve estimation accuracy, without risk of sub-optimal convergence, as all other homogeneous updates to the posteriors are applied and derived from a stationary and identical distribution of ground-truth information.

\paragraph{Proof:} Assume $T$ be an ergodic state distribution represented by a Dirac delta function $\delta(z)$, with its differential entropy of $T$ approaching $-\infty$. Let $X, Y$ be two independent distributions that are identically distributed with respects to $T$. Hence, if the relational entropy is defined as $\mathcal{H}(X) \leq \mathcal{H}(Y)$, then we assert the relative cross entropy with respects to $T$ to be $H(T,X) \leq H(T,Y)$, or equivalently, we state $P_X(z) \geq P_Y(z)$, under the assumption of the only updates that were made is restricted to the ones we defined. In other words, $X$ is covered by $Y$, and $T$ is covered by $X$.

Let $D(P,Q) \propto \mathcal{H}(Q,P)$ be a divergence measure that is proportional to the cross entropy. Following this definition, if $\mathcal{H}(X) \leq \mathcal{H}(Y)$, then we can assert $D(T,X) \leq D(T,Y)$ assuming no other conflicting terms dominating the measure. Examples of divergence measure that satisfies this property, either by definition (see Equation \ref{kl_proof}) or by lower bound, include reverse KL divergences and Wasserstein distance \cite{entropy}.
\begin{equation} \label{kl_proof}
    D_{KL}(T|X) - D_{KL}(T|X) = H(T,X) - H(T,Y) \leq 0
\end{equation}
Alternatively, we can intuitively reason that estimates with lower entropy reflect a more recency in non-masked update and thereby hold a greater information gain. This would result in a greater overlap to the target distribution.

\paragraph{External beliefs.} With state estimation of external entities, we follow the same procedure described above, however, we ensure there is no information leakage between non-cooperative agents. So, we apply the same masking for both control and observation updates, which are both curated for the task at hand.

Additionally, we ensure that there is no global information leakage by using the beliefs and information provided at the local level. So, for the observation update in Predator-Prey task, we determine whether the indicator bit $I$ is set to figure out if the prey is within the sensing range. We use the agent estimation to attempt to localize itself and propagate the prey's estimate accordingly to the sensing range using a tensor operator $\mathcal{T}_s$. This operator is defined similarly as the transition dynamics, but according to the sensing range. Then, we process the update to the corresponding external agent's belief $e$.

\begin{equation}
\textit{bel}_{(i,e),t} = \eta (\mathcal{T}_s \textit{bel}_{(i,i),t} \odot  \textit{bel}_{(i,e),t})
\end{equation}

Lastly, we note that the uncertainty of belief distributions is strongly coupled to the decision-making process. We limit our assertions within the scope of pure state estimation while acknowledging that control does play a large role into fully optimizing its performance (i.e. balancing exploration).

\section{Experiments}
Our experiments provide insight into whether our modifications sufficiently accommodate for the limitations of decentralized communication networks and the impacts of the different design considerations we impart within the filtering system on both tasks.

\subsection{Setup}
\begin{figure}

    \centering
    \begin{subfigure}{\textwidth}
        \centering
        \includegraphics[width=\textwidth]{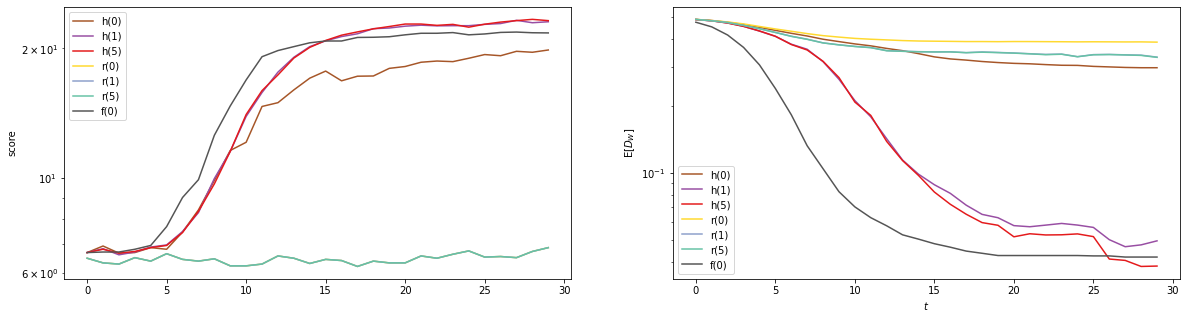}
        \caption{Congregation task.}
    \label{fig:congregation}
    \end{subfigure}
    \centering
    \begin{subfigure}{\textwidth}
        \centering
        \includegraphics[width=\textwidth]{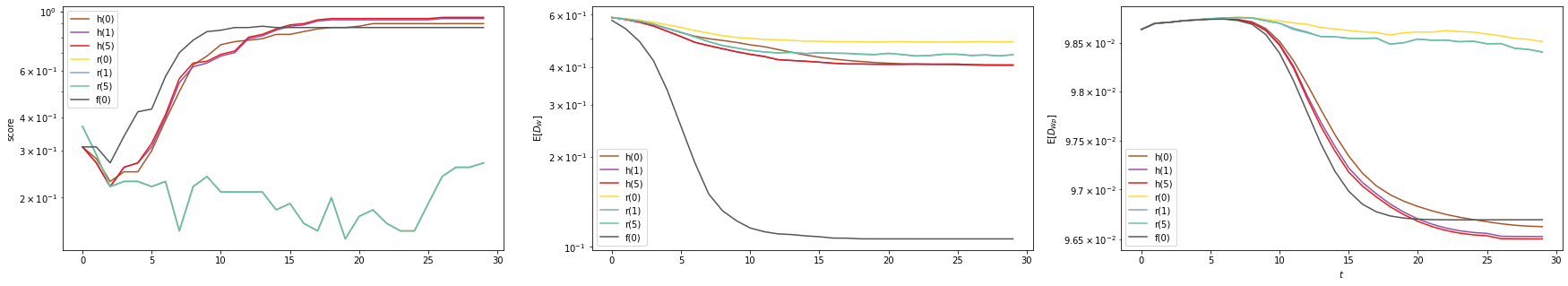}
        \caption{Predator-Prey task.}
    \label{fig:predatorprey}
    \end{subfigure}
    \caption{Quantitative evaluations of Dec-MAS filtering systems: The average is shown, which is taken from $100$ simulation runs of the cumulative reward (score) and total expected $1-$Wasserstein distance $\mathrm{E}[D_w]$ and the expected $1-$Wasserstein distance of the prey $\mathrm{E}[D_{wp}]$ from the target distribution for Figure \ref{fig:predatorprey}. The decentralized system $h(\cdot)$, the oracle system $f(\cdot)$, and the random system $r(\cdot)$  are simulated, and the number of message passing rounds tested are $0,1,5$ for the random and heuristic policies.}
\end{figure}

To evaluate our system, we use the same configurations described for the two tasks using the respective control policies from Section \ref{system_overview}. We set the size of the gridworld to be $H = W = 10$, and the number of agents to be $5$, each with a sensing range of $1$. As our metric, we track the cumulative reward, which acts as a surrogate measure for information consistency between agents' beliefs, and the expected $1-$Wasserstein distance between the local belief distributions to the true distribution (represented using a Dirac delta function), which gives a direct measure of true divergence. All experiments are ran on $100$ different random seeds on environments based on \cite{1606.01540, magym}. The results can be seen in Figures \ref{fig:congregation} and \ref{fig:predatorprey}.

\subsection{Using random control policies. } We first measure the impact of the choice of control policies, comparing random and heuristic policies using our decentralized filtering system. As the random policies do not act according to their knowledge base, the actual performance does not reveal much information other the fact that these tasks cannot be solved using random controls. The total divergence between the true distribution does show a large difference between the estimation accuracy of the heuristic and random policy. This result reaffirms the significance of the control policy in terms of true divergence of state estimation, as each latent state holds a certain amount of information gain and the controls largely dictates what states are observed.

\subsection{Comparing against oracle system. } The oracle system is a fully-connected communication network, where the sensing range of each agent is defined to encompass the entire map. We define the oracle system to follow the heuristic policies and to use classical Bayes filtering without any message passing and belief sharing. The reward function for both tasks still require the agents to be within same sensing range as the decentralized systems. The results show that the oracle system does outperform our decentralized systems on both tasks initially, which was as expected since every agents would be exposed to updates with the highest possible information gain. This assertion is most evident when measuring the total true divergence from the Predator-Prey experiments. However, interestingly, the performance of our systems eventually surpasses the oracle, converging at a higher score and lower true divergence. Hence, we demonstrate that even with limited decentralized communication, the local belief distributions can still converge near the true distribution with proper modifications.

As to reason why our system outperforms the oracle system, we argue that the control policy employed does not necessarily balance the exploration-exploitation needed to perform well in these tasks. The greater uncertainty introduced by limited information gain provided the needed exploration, potentially leading to lower true divergence.

% Ultimately, we conclude that our modified Bayes filtering algorithm applied on partially-observable Dec-MAS sufficiently sufficiently accommodates for the limited sensing range of decentralized communication.

\subsection{Impact of increasing message passing rounds. }
When we observe systems with more message passing rounds, the improvement is mostly apparent in the true divergence, while reflecting similar patterns in the performances. Even across the random policies, the true divergence is lower for systems with message passing. The results do reflect that there exists a fairly significant diminishing return on increasing the number of message passing rounds, with a steep drop in improvement from $\geq1$ message passing rounds. Perhaps in larger and dense environments, greater transitive communication would be more significant. 

\section{Conclusion}
In this work, we proposed extensions to the Bayes filter algorithm to pair better with multi-agent systems using decentralized communication networks. Our adjustments included a greedy entropy-based belief sharing step and modifications to the environment model by defining masked updates. Our results support that our modified Bayes filtering algorithm applied on Dec-MAS sufficiently accommodates for the limited sensing range of decentralized communication. In future works, efforts can be focused on devising control policies using the feature space developed in this work.

% \bibliographystyle{mla}
% \bibliography{jfrExampleRefs}
\printbibliography

\end{document}